\documentclass[aps,prl,preprint,showpacs]{revtex4}
\usepackage{bm}
\usepackage{graphicx}
\begin{document}
\title{Energy partition and segregation for an intruder
in a vibrated granular system under gravity}
\author{J. Javier Brey}
\email{brey@us.es}
\author{M.J. Ruiz-Montero}
\author{F. Moreno}
\affiliation{F\'{\i}sica Te\'{o}rica, Universidad de Sevilla,
Apdo.\ de
 Correos 1065, E-41080 Sevilla, Spain}

\date{today}

\begin{abstract}
The difference of temperatures between an impurity and the
surrounding gas in an open vibrated granular system is studied. It
is shown that, in spite of the high inhomogeneity of the state,
the temperature ratio remains constant in the bulk of the system.
The lack of energy equipartition is associated to the change of
sign of the pressure diffusion coefficient for the impurity at
certain values of the parameters of the system, leading to a
segregation criterium. The theoretical predictions are consistent
with previous experimental results, and also in agreement with
molecular dynamics simulation results reported in this paper.
\end{abstract}


\maketitle

Granular systems, i.e. assemblies of macroscopic grains
dissipating part of their kinetic energy during collisions,
exhibit a rich phenomenology with many qualitative differences as
compared with molecular systems \cite{JNyB96}. Among these are the
breakdown of energy equipartition \cite{WyP02,FyM02} and particle
segregation in agitated mixtures \cite{DRyC93,BEKyR03}. The aim of
this work is to investigate both effects for a dilute mixture in
the tracer limit, i.e. very low mole fraction of one of the
species, showing that there is a close relationship between them.

Different species of a granular mixture have different granular
temperatures, defined from the average kinetic energy of the
particles. A prediction for their ratio in a dilute binary mixture
of smooth inelastic hard spheres has been derived from the
inelastic Boltzmann equations describing the time evolution of the
distribution functions of the species \cite{GyD99}, and its
accuracy has been verified by Molecular Dynamics simulations for
the particular case of an isolated homogeneous system
\cite{DHGyD02}. In this Letter, it will be shown that kinetic
theory also accurately predicts the temperature ratio in the case
of a highly inhomogeneous driven system. This ratio turns out to
be constant in the bulk of the system, consistently with the
experimental results reported in \cite{FyM02}. As in the
homogeneous state, deviations from equipartition depend on the
mechanical differences between the species and the degrees of
inelasticity of collisions.

Particle segregation is the demix of a granular mixture when
shaken. Usually, the larger particles are observed to rise
(Brazil-nut effect), although under certain conditions they can
also tend to descend (reverse Brazil-nut effect). Several
mechanisms, corresponding to different scenarios, have been
proposed to explain both behaviors \cite{DRyC93,HQyL01,JyY02},
although the phenomenon is far from being fully understood. Here,
segregation will be investigated in the context of hydrodynamics
for a dilute granular mixture as derived from kinetic theory
\cite{GyD02}. The relative position of the tracer component with
respect to the excess component is determined by the sign of the
pressure diffusion coefficient. While in a molecular gas this sign
is fixed by the mass ratio of the particles of the components, for
a granular gas it also depends on the temperature ratio. Due to
the lack of energy equipartition, if follows that the criterion
for segregation is rather complicated, involving all the
parameters of the mixture.

The model system considered is a low density gas of smooth
inelastic hard spheres ($d=3$) or disks ($d=2$) of mass $m$ and
diameter $\sigma$, and one impurity of mass $m_{0}$ and diameter
$\sigma_{0}$. This is formally equivalent to the tracer limit for
the impurity component. Inelasticity of collisions is specified by
two independent constant coefficients of normal restitution,
$\alpha$ and $\alpha_{0}$, referring to gas-gas and impurity-gas
collisions, respectively. The system is in presence of a uniform
external field of the gravitational type, so each particle is
submitted to a force per unit of mass given by $-g_{0}
\widehat{\bm e}_{z}$, where $g_{0}$ is a positive constant and
$\widehat{\bm e}_{z}$ a unit vector in the direction of the $z$
axis. Energy is continuously supplied to the system through the
bottom wall located at $z=0$ that is vibrating with small
amplitude and high frequency. There is no upper wall, i.e. the
system is open.

Under the above conditions, the system exhibits an inhomogeneous
steady state with gradients only in the $z$ direction and
vanishing velocity field. In the case of a one-component system,
it has been verified that the hydrodynamic profiles away from the
walls are well described by the hydrodynamic equations
\cite{BRyM01,ByR04a}. It is assumed that these profiles are not
affected by the inclusion of the impurity. Consider the local
temperatures of the gas and the impurity, $T(z)$ and $T_{0}(z)$,
defined in the usual way from the respective mean square
velocities (with the Boltzmann constant set equal to unity). A
formal relation between both temperatures can be derived if the
existence of a hydrodynamic regime is assumed and, moreover, that
the corresponding ``normal'' solution of the Boltzmann equation
for the impurity can be generated by an extension of the
Chapman-Enskog method \cite{GyD02,BRyM05b}. Then, it is obtained
that the lowest order in the gradients of the cooling rates for
the gas, $\zeta^{(0)}(z)$, and the impurity, $\zeta_{0}^{(0)}(z)$,
must be equal. These rates are nonlinear functionals of the zeroth
order distribution functions, and can be estimated at good
approximation using Maxwellians, with the results \cite{DByL02}
\begin{equation}
\label{1} \zeta^{(0)*} \equiv \frac{\zeta^{(0)}(z)}{n(z)
\sigma^{d-1} v_{g}(z)} =\frac{\sqrt{2} \pi ^{(d-1)/2}}{\Gamma
(d/2)d}\, (1-\alpha^{2}),
\end{equation}
\begin{equation}
\label{2} \zeta_{0}^{(0)*} \equiv \frac{\zeta_{0}^{(0)}(z)}{n(z)
\sigma^{d-1} v_{g}(z)} = \nu^{*}_{0} (1+\phi)^{1/2}
\left(1-h\frac{1+\phi}{\phi} \right),
\end{equation}
where $n(z)$ is the number density of the gas,
$v_{g}(z)=[2T(z)/m]^{1/2}$, $h=m(1+\alpha_{0})/2(m+m_{0})$,
$\nu^{*}_{0}$ is a dimensionless impurity-particle collision rate,
\begin{equation}
\label{3} \nu^{*}_{0}=\frac{8h \pi^{(d-1)/2}}{\Gamma(d/2)d} \left(
\frac{\overline{\sigma}}{\sigma} \right)^{d-1},
\end{equation}
with $\overline{\sigma}=(\sigma + \sigma_{0})/2$, and $\phi$ the
ratio of mean square velocities,
\begin{equation}
\label{4} \phi=\frac{ m T_{0}(z)}{m_{0} T(z)}.
\end{equation}
Equating Eqs. (\ref{1}) and (\ref{2}) provides the equation for
$\phi$,
\begin{equation}
\label{5} (1+\phi)^{1/2} \left( 1-h \frac{1+\phi}{\phi} \right) =
\frac{\beta}{h},
\end{equation}
\begin{equation}
\label{6} \beta \equiv  \frac{1-\alpha^{2}}{4 \sqrt{2}} \left(
\frac{\sigma}{\overline{\sigma}} \right)^{d-1}.
\end{equation}
This gives a cubic equation which has a unique real, positive
solution for all the allowed values of $h$ and $\beta$. For
elastic collisions one gets $\phi= m/m_{0}$, as required by energy
equipartition. Moreover, in the limit of an elastic gas,
$\alpha=1$, but inelastic collisions between the intruder and the
fluid, $\alpha_{0}<1$, the expression obtained in ref.
\cite{MyP99} is recovered. The behavior of the solution in the
limit $\phi \rightarrow 0$ in the context of the isolated
homogeneous state has been analyzed in \cite{SyD01}, where it has
been shown that a qualitative change similar to a second order
phase transition occurs, but this will not be discussed here.

From Eq.\ (\ref{5}) it follows that $\phi$ does not depend on $z$.
Therefore, the temperature ratio $T_{0}(z)/T(z)$ not only differs
in general from unity, but  remains constant along the system.
Moreover, it does not depend either on the precise way in which
the wall is being vibrated, as long as the assumed steady state is
reached. Of course, the result only holds in the bulk region of
the system, where a hydrodynamic description is expected to apply.
These predictions are in qualitative agreement with the
experimental results found by Feitosa and Menon \cite{FyM02},
although it must be realized that their experiments are not
carried out in the tracer limit. On the other hand, these authors
did not find any dependence of the temperature ratio on the
inelasticity of the grains, while Eq.\ (\ref{5}) predicts a rather
strong one.

\begin{figure}
\includegraphics[scale=0.5,angle=-90]{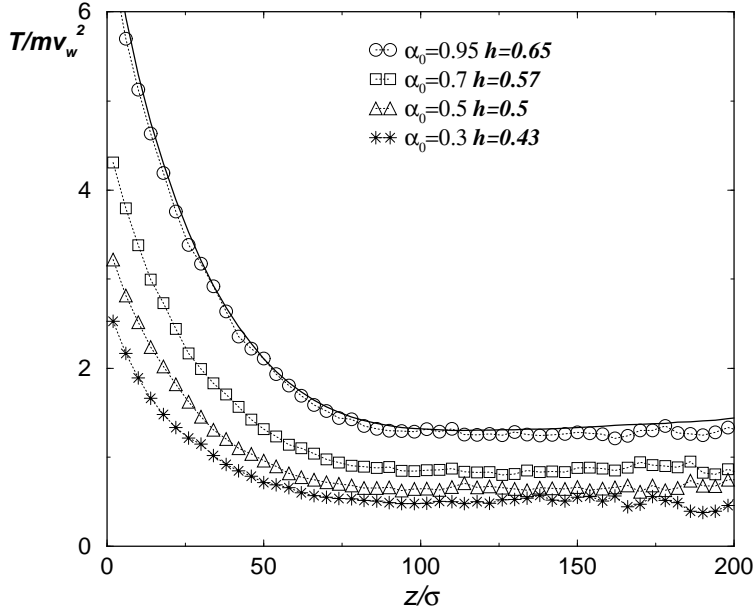}
\caption{Dimensionless temperature profiles of the gas (solid
line) and the impurity (symbols) for $m_{0}/m=1/2$. The different
symbols correspond to different values of the restitution
coefficient $\alpha_{0}$, as indicated. The values of the
parameter $h$, defined in the main text, are also indicated for
reference. \label{fig1}}
\end{figure}

\begin{figure}
\includegraphics[scale=0.5,angle=-90]{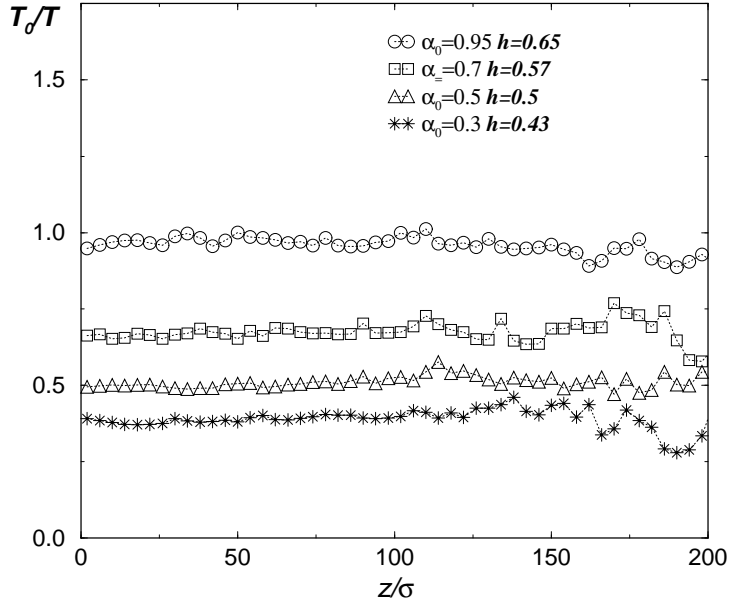}
\caption{Profiles of the ratio between the impurity and the gas
temperature for the same values of the parameters as in Fig.
\ref{fig1}. \label{fig2}}
\end{figure}

To check the above results, we have performed molecular dynamics
(MD) simulations of a system of $N$ inelastic hard disks ($d=2$).
The wall at the bottom vibrates with a sawtooth profile with
velocity $v_{w}$ and collisions of the particles with it are
elastic. Moreover, the amplitude of the wall motion is much
smaller than the mean free path of the particles next to it, so in
practice the position of the wall can be taken as fixed at $z=0$.
Periodic boundary conditions are used in the direction
perpendicular to the field. The units are defined by
$m=\sigma=g_{0}=1$. In all the results to be reported in the
following, the parameters determining the state of the gas have
been kept constant, namely $N=359$, $\alpha=0.95$, $v_{w}=5$, and
a width $S=50$. For these values, it is verified that the system
is fluidized and actually reaches the steady state assumed above.
In Fig.\ \ref{fig1}, the temperatures profiles  of the impurity
for $m_{0}/m=1/2$, $\sigma_{0}=1$, and several values of
$\alpha_{0}$ are shown. Also, the temperature profile of the gas
(solid line) is included. It is clearly observed that energy
equipartition is not verified, and that the temperature difference
depends on $\alpha_{0}$. The ratio between the impurity and gas
temperatures for the same states is plotted in Fig. \ref{fig2}. In
agreement with the theoretical prediction, the ratio remains
constant, aside from statistical fluctuations, in the interior of
the system, even in the region where each of the partial
temperatures presents large gradients. In fact, the kinetic
boundary layer next to the vibrating wall is quite narrow. Similar
behaviors have been found for other values of $m_{0}$, namely
$m_{0}/m=0.75,1$ and $2$.

\begin{figure}
\includegraphics[scale=0.5,angle=-90]{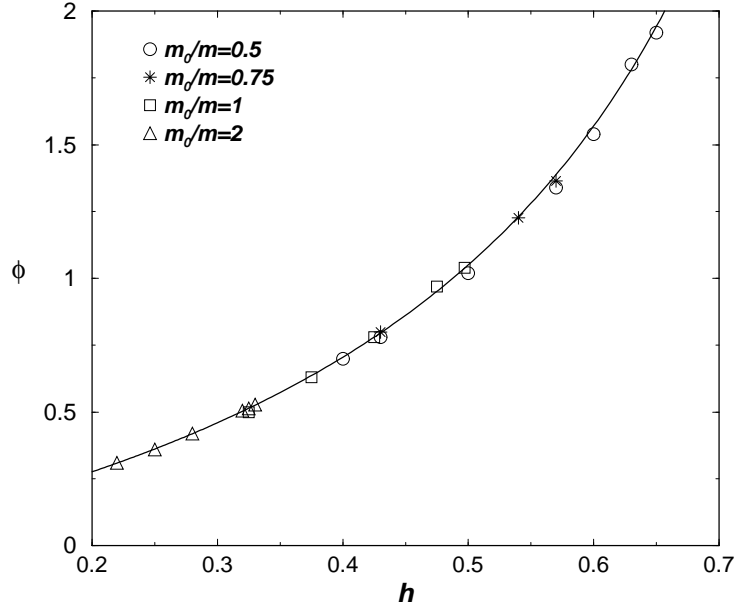}
\caption{Ratio of the mean square velocities $\phi$ as a function
of the parameter $h$ defined in the main text. The solid line is
the theoretical prediction given by Eq.\ (\ref{5}) and the symbols
are from the simulations for different values of the mass ratio,
as indicated . \label{fig3}}
\end{figure}

The quantitative comparison of the simulation results with Eq.\
(\ref{5}) is presented in Fig.\ \ref{fig3}, where $\phi$ is
plotted against the parameter $h$. The solid line is the solution
of the equation and the different symbols correspond to different
values of the mass ratio as indicated. A very good agreement is
obtained. It is worth to stress that the theory remains accurate
up to very small values of the coefficient of restitution
$\alpha_{0}$, indicating the validity of the kinetic and
hydrodynamic descriptions even for very strong dissipation. Also,
for later discussion note that $\phi$ can be larger than unity
even if $m/m_{0} \leq 1$, and viceversa.

Consider next the number density for the impurity in the steady
state, $n_{0}(z)$. It is determined by the condition that the
associated flux $j_{z}$ must vanish. To first order in the
gradients, it is
\begin{equation}
\label{7} j_{z}= -m_{0} D \partial_{z} x_{0} -
\frac{mn}{T(z)}D^{\prime}
\partial_{z}T -\frac{m}{T(z)}D_{p}
\partial_{z} p(z),
\end{equation}
where $x_{0}=n_{0}/n$, $p=nT$ is the pressure, $D$ the diffusion
coefficient, $D^{\prime}$ the thermal diffusion coefficient, and
$D_{p}$ the pressure diffusion coefficient. Explicit expressions
for these transport coefficient can be derived  by the
Chapman-Enskog procedure in the first Sonine approximation
\cite{GyD02,BRyM05b},
\[
D=\frac{nT_{0}}{m_{0}} \left(\nu-\frac{\zeta^{(0)}}{2}
\right)^{-1}, \quad D^{\prime}=- \frac{\zeta^{(0)}}{2 \nu}D_{p},
\]
\begin{equation}
\label{8} D_{p}=\frac{n_{0}T_{0}}{mn} \frac{\phi-1}{\phi} \left(
\nu -\frac{3 \zeta^{(0)}}{2}+\frac{ \zeta^{(0)2}}{2 \nu}
\right)^{-1}.
\end{equation}
Here $\nu$ is a collision frequency,
\begin{equation}
\label{9} \nu=\nu_{e} \frac{1+\alpha}{2} \left(1-\Delta
\right)^{1/2} \left(1+\phi \right)^{1/2},
\end{equation}
where $\Delta=m/(m+m_{0})$ and $\nu_{e}$ is the elastic limit,
\begin{equation}
\label{10} \nu_{e}=\frac{4 \sqrt{2} \pi^{(d-1)/2}}{\Gamma (d/2) d}
\overline{\sigma}^{d-1} \Delta^{1/2} n \left( \frac{T}{m_{0}}
\right)^{1/2}.
\end{equation}
Although the equation $j_{z}=0$ can be (numerically) integrated to
get $n_{0}(z)$, given that the hydrodynamic profiles for the gas
are known,  we will restrict ourselves to a qualitative property
of $n_{0}(z)$. We will determine the position of the maximum of
$n_{0}(z)$ relative to that of $n(z)$. That means that we consider
situations in which the latter exists, which is always the case
for large enough number of particles $N$ \cite{BRyM01}. Let us
denote by $z_{m}$ the height at which this maximum occurs. It has
been established that at this point the temperature of the gas is
a decreasing function of $z$, i.e. $\partial T /\partial z <0$ at
$z=z_{m}$ \cite{BRyM01}. Then, it is obtained
\begin{eqnarray}
\label{11} \left( \frac{\partial \ln n_{0}}{\partial z}
\right)_{z_{m}} & = & \frac{\phi -1}{\phi} \left( 1-
\frac{\zeta^{(0)}}{2 \nu} \right)^{2} \nonumber \\
 & & \times \left(1- \frac{3
\zeta^{(0)}}{2 \nu} +\frac{\zeta^{(0)2}}{2 \nu^{2}} \right)^{-1}
\left| \frac{\partial \ln T}{\partial z} \right|_{z_{m}}.
\end{eqnarray}
If the right hand side of the above equation is positive,
$n_{0}(z)$ is still growing at $z=z_{m}$, following that its
maximum occurs at $z_{0,m} >z_{m}$. In the opposite case, the
maximum of $n_{0}(z)$ takes place at $z_{0,m}<z_{m}$.  Note that
this sign is the same as that of the pressure diffusion
coefficient $D_{p}$. The general discussion is rather complicated,
given the large number of parameters involved \cite{BRyM05b}. For
the sake of simplicity, we are going to consider the region of the
parameter space verifying $\beta < h/2$, which includes all the MD
simulations reported here. For this range of values, it is easily
verified that $1-3\zeta^{(0)}/ 2 \nu+\zeta^{(0)2}/2 \nu^{2}
>0$, and the sign of $\partial \ln n_{0} /\partial z$ at
$z=z_{m}$ is determined by the value of $\phi$. For $\phi >1$
($\phi <1$) the derivative is positive (negative) and the position
of the density maximum for the impurity is higher (lower) than
that for the gas. As mentioned above, the value of $\phi$ can be
quite different from the mass ratio due to the different
temperatures of the impurity and the gas. Since the value of
$\phi$ is a function of all the parameters of the system, the
segregation criterion, understood as a rule determining the
relative position of the impurity with respect to the gas, also
involves all those parameters.

\begin{figure}
\includegraphics[scale=0.5,angle=-90]{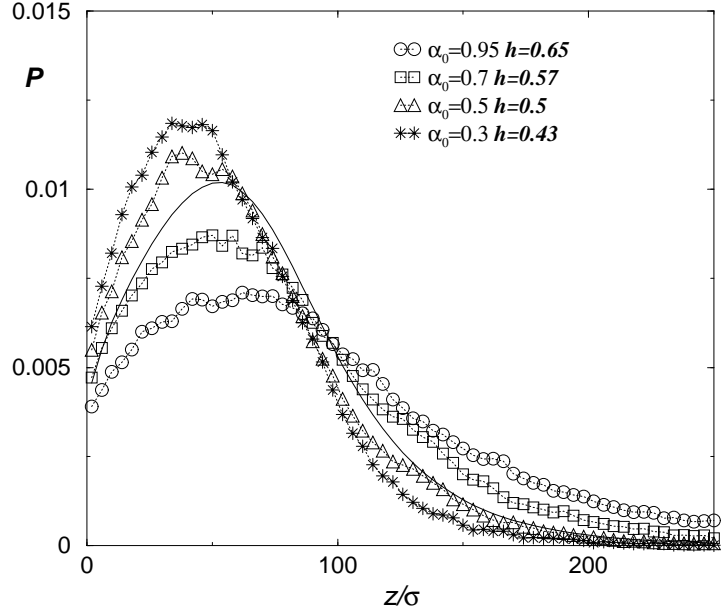}
\caption{Normalized height distributions for the gas (solid line)
and the impurity (symbols) for the same systems as in Fig.\
\ref{fig1}. \label{fig4}}
\end{figure}

The MD density profiles for the same systems as in Fig.\
\ref{fig1} are given in Fig.\ \ref{fig4}. Again, the solid line is
the density profile of the gas, normalized to unity for comparison
purposes. As the restitution coefficient $\alpha_{0}$ increases,
the maximum of the density profile for the impurity moves towards
higher regions and the profile becomes wider, showing the tendency
of the impurity to rise. The same effect is observed for the other
values of the mass ratio considered. Since the value of $\phi$
increases as $\alpha_{0}$ increases keeping the remaining
parameters constant, this behavior is in qualitative agreement
with the above theoretical predictions.

\begin{figure}
\includegraphics[scale=0.5,angle=-90]{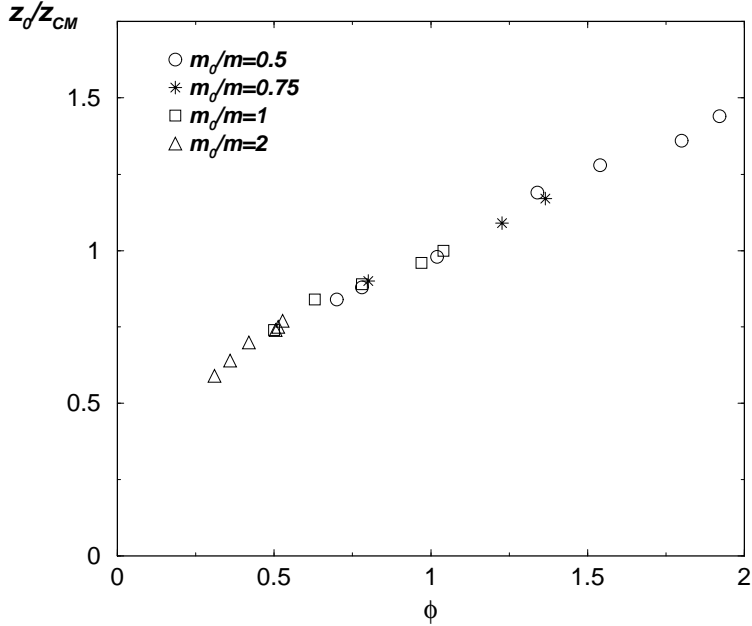}
\caption{Ratio between the impurity and the gas center of mass
positions versus the ratio of the mean square velocities $\phi$.
The different symbols correspond to different values of the mass
ratio as indicated. \label{fig5}}
\end{figure}

The distributions for the position of the impurity become quite
flat as the value of $\alpha_{0}$ increases, rendering quite
difficult to identify their maximum from the MD results. Then,
what has been measured is the center of mass position for both the
gas, $z_{cm}$, and the impurity, $z_{0}$. In Fig.\ \ref{fig5} the
ratio $z_{0}/z_{cm}$ versus $\phi$ is shown. It is seen that the
segregation criterium derived above is fairly obeyed, in spite of
the fact that the shape of the density distribution clearly
indicates that the position of the maxima are rather different
from those of the centers of mass. Similar results have been found
for other values of the parameters, including different sizes of
the impurity \cite{BRyM05b}.

To conclude, it must be stressed that the present work deals with
a dilute granular mixture in the tracer limit. At higher
densities, other segregation mechanisms, such as those discussed
in \cite{HQyL01} and \cite{JyY02}, become important. On the hand,
there is no reason to expect that the one identified here fails to
be relevant at those densities.

This research was supported by the Ministerio de Educaci\'{o}n y
Ciencia (Spain) through Grant No. FIS2005-01398 (partially
financed by FEDER funds).

\end{document}